\documentclass[superscriptaddress,twocolumn,showpacs,preprintnumbers,amsmath,amssymb,floatfix,nofootinbib]{revtex4-1}

\usepackage{color}   
\usepackage{graphicx}
\usepackage{dcolumn}

\begin{document}

\def\bb    #1{\hbox{\boldmath${#1}$}}

\title{ From QCD-based hard-scattering
to
 nonextensive statistical mechanical descriptions \\ of
 transverse momentum spectra in high-energy $pp$ and $p\bar p$ collisions
}

\author{Cheuk-Yin Wong}
\email{wongc@ornl.gov}
\affiliation{Physics Division, Oak Ridge National Laboratory,
Oak Ridge, Tennessee 37831, USA}
\author{Grzegorz Wilk}
\email{wilk@fuw.edu.pl}
\affiliation{ National Centre for Nuclear Research, Warsaw 00-681, Poland}
\author{ Leonardo J.\ L.\ Cirto}
\email{cirto@cbpf.br}
\affiliation{
Centro Brasileiro de
Pesquisas Fisicas \& National Institute of Science and Technology
for Complex Systems,\\ \hspace*{0.02cm}~~Rua Xavier Sigaud 150, 22290-180 Rio de
Janeiro-RJ, Brazil }
\author{ Constantino Tsallis}
\email{tsallis@cbpf.br}
\affiliation{
Centro Brasileiro de
Pesquisas Fisicas \& National Institute of Science and Technology
for Complex Systems,\\ \hspace*{0.02cm}~~Rua Xavier Sigaud 150, 22290-180 Rio de
Janeiro-RJ, Brazil }
\affiliation{
Santa Fe Institute, 1399 Hyde Park Road,
Santa Fe, NM 87501, USA}

\begin{abstract}
Transverse spectra of both jets and hadrons obtained in high-energy
$pp$ and $p\bar p $ collisions at central rapidity exhibit power-law
behavior of $1/p_T^n$ at high $p_T$.  The power index $n$ is 4-5 for
jet production and is 6-10 for hadron production.
Furthermore, the hadron spectra spanning over 14 orders of magnitude
down to the lowest $p_T$ region in $pp$ collisions at LHC can be
adequately described by a single nonextensive statistical mechanical
distribution that is widely used in other branches of science.  This
suggests indirectly the possible dominance of the hard-scattering process over
essentially the whole $p_T$ region at central rapidity in high energy $pp$
and $p \bar p$ collisions.  We show here direct evidences of such a dominance
of the hard-scattering process by investigating the power indices of UA1
and ATLAS jet spectra over an extended $p_T$ region and the two-particle
correlation data of the STAR and PHENIX Collaborations in high-energy
$pp$ and $p \bar p$ collisions at central rapidity.  We then study how
the showering of the hard-scattering product partons alters the power
index of the hadron spectra and leads to a hadron distribution that
may be cast into a single-particle nonextensive statistical mechanical
distribution. Because of such a connection, the nonextensive
statistical mechanical distribution may be considered as a
lowest-order approximation of the hard-scattering of partons followed
by the subsequent process of parton showering that turns the jets into
hadrons, in high energy $pp$ and $p\bar p$ collisions.

\end{abstract}


\maketitle
\section{Introduction}
\label{Introduction}

Transverse momentum distribution of jets and hadrons provide useful
information on the collision mechanisms and their subsequent
dynamics. The transverse spectra of jets in high-energy $pp$ and
$p\bar p$ experiments at high $p_T$ and central rapidity exhibit a
power-law behavior of $1/p_T^n$ with the power index $n$ $\sim$ 4 - 5,
which indicates that jets are scattered partons produced in
relativistic hard-scattering processes
\cite{Bla74,Ang78,Fey78,Owe78,Duk84,Sj87,Wan91,Won94,Arl10,TR13,Won12,Won13Is,Won13}.
On the other hand, the power index  for hadron spectra are in the range of 6  to 10, slightly
greater than those for jets
\cite{Ang78,TR13,Arl10,Won12,Won13Is,Won13}, revealing that hadrons
are showering products from jets, and the hadron spectra are modified
from the jet spectra but retaining the basic power-law structure of
the jet spectra.  It was found
\cite{Won12,Won13,Won13Is,Won14EPJ,CTWW14} recently that the hadron
spectra spanning over 14 decades of magnitude from the lowest to the
highest $p_T$ at central rapidity can be adequately described by a
single nonextensive statistical mechanical distribution that is widely
used in other branches of sciences \cite{T1,Tsa14},
\begin{equation}
F\left(p_{T}\right)=A~\left[1-\left(1-q\right)\frac{p_{T}}{T}\right]^{1/(1-q)}.
\label{T}
\end{equation}
Such a distribution with $q=1+1/n$ is phenomenologically equivalent to the
quasi-power law interpolating formula introduced by Hagedorn \cite{H}
and others \cite{Michael}
\begin{eqnarray}
F(p_T)=A \left ( 1 + \frac{p_T}{p_0} \right ) ^ {-n}, \label{CM-H}
\end{eqnarray}
for relativistic hard scattering. Both Eqs. (\ref{T}) and
(\ref{CM-H}) have been widely used in the phenomenological
analysis of multiparticle productions, cf., for example,
\cite{BCM,Beck,RWW,B_et_all,JCleymans,ADeppman,Others,WalRaf,WWreviews}
and references therein.

It is of interest to know why such a nonextensive statistical
mechanical distribution (\ref{T}) may be a useful concept for hadron
production.  It may also be useful to contemplate its possible
implications.  The shape of the spectrum reflects the complexity, or
conversely the simplicity, of the underlying production mechanisms.
If there are additional significant contributions from other
mechanisms, the specification of the spectrum will require degrees of
freedom additional to those of the relativistic hard-scattering model.
The small number of apparent degrees of freedom
 of the spectrum over
such a large $p_T$ region\footnote{
There are only three degrees of freedom in Eq.\ (1):  $A$, $q$ (or equivalently $n$), and $T$.   Notice that three degrees of freedom  are almost the minimum number to specify a spectrum.  Our spectrum is  therefore very  simple.  
It is interesting therefore  that  the counting of the degrees of freedom in our case can lead to the suggestion of possible hard-scattering dominance and the successful search for supporting direct evidences as we describe in the present work.
} 
suggests the possible dominance of the
hard-scattering process over essentially  the whole region of $p_T$ at central
rapidity \cite{Won12,Won13,Won13Is,Won14EPJ}.

The counting of the degrees of
freedom provides merely an indirect evidence for the dominance of the hard-scattering process over the whole $p_T$ region.  We would like to search for
direct evidences for such a dominance in three different ways.  The hard
scattering process is characterized by the production of jets whose
transverse spectra carries the
signature of the power index of $n$ $\sim$ 4 - 5 at central rapidity
\cite{Won13,Won13Is,Won14EPJ}.  The relevant data come from
well-defined jets with transverse momenta greater than many tens of
GeV obtained in D0, ALICE, and CMS Collaborations
\cite{Abb01,Alice13,cms11}.  To seek direct supporting evidences of
jet production by the hard-scattering process over the lower-$p_T$
region, we examine the experimental UA1 and ATLAS data which give the
invariant cross sections for the production of jets from the low-$p_T$ region (of a few GeV) to the high-$p_T$
region (up to 150 GeV) \cite{UA188,Aad11}.  If the power index $n$ of the
UA1 and ATLAS jet spectra  at central rapidity is
close to 4 - 5, it will constitute a direct evidence of the dominance of
the hard-scattering process over the extended $p_T$ region, for $p\bar
p$ and $pp$ collisions at high energies.  In such an analysis, we need to take
into account important $p_T$-dependencies of the structure function
and the running coupling constant by refining the analytical formula
of the hard-scattering integral.

The hard-scattering process is characterized by the production of jets as angular clusters of hadrons.
We can seek additional direct evidences for
the dominance of the hard-scattering process by searching for hadron angular clusters 
 on the near-side
using the  two-particle correlation data in high-energy $pp$ collisions from STAR and PHENIX
Collaborations
\cite{STAR05,STAR06,Put07,PHEN08,STAR06twopar,Por05,Won08,Won09,Tra11,Ray11}.
Two-particle angular correlation data are specified by the azimuthal angular difference $\Delta \phi$ and the pseudorapidity difference $\Delta \eta$ of the two particles.
If hadrons associated 
with a low-  and high-$p_T$ trigger are correlated 
at $(\Delta \phi, \Delta \eta)\sim 0$ on the near-side, it will constitute 
an indication of the dominance of  the hard-scattering process over essentially  the
whole  $p_T$ region.

Finally, the
hard-scattering process is characterized by the production of two jets of particles.
We can seek an additional direct evidence for the other partner jet
by searching for angular clusters of associated hadrons
 on the away-side in two-particle correlation data from STAR and PHENIX
Collaborations
\cite{STAR05,STAR06,Put07,PHEN08,STAR06twopar,Por05,Won08,Won09,Tra11,Ray11}.
A ridge of hadrons on the away-side at $\Delta \phi \sim \pi$ associated with a low-$p_T$ and 
high-$p_T$ trigger will  indicate the production of the   partner jet   by the hard-scattering process over the
whole  $p_T$ region.

While our search has been stimulated by the simplicity of the hadron $p_T$  spectrum, it should be mentioned that the importance of the production of jets with $p_T$ of a few GeV (minijets) has already been well emphasized in the earlier work of \cite{Wan91} ,and the production of the low-$p_T$ jet (minijets) in the low-$p_T$ region has been pointed in the work of \cite{ STAR06twopar,Por05,Tra11,Ray11}. 
We are seeking here   a synthesizing  description linking these advances together into 
a single and simplifying observation on the dominance of the hard-scattering over the whole $p_T$ region,  with a special  emphasis on the production mechanism.
Such a complementary and synthesizing viewpoint may serve the useful purposes of helping guide our intuition and summarizing important features of the collision process.

After examining the experimental evidences for  the dominance of 
the relativistic hard-scattering process  in the whole
$p_T$ region, we would like to understand how jets turn into hadrons
and in what way the jet spectra evolves to become the hadron spectra by
parton showering.  Our understanding may allow us to bridge the
connection between the hard-scattering process for jet production and
its approximate representation by a nonextensive statistical
mechanical distribution for hadron production.  In consequence, the
dominance of the hard-scattering process over the whole
$p_T$ region may allow the nonextensive statistical mechanical
distribution to describe the observed hadron transverse spectra
spanning  the whole $p_T$ region at central rapidity, in $pp$
collisions at LHC.

In this paper, we restrict our attention to the central rapidity
region at $\eta\sim 0$ and organize the paper as follows.  In Section
II, we review and refine the analytical results for the relativistic
hard-scattering process.  We use the analytical results to analyze the
spectra for high-$p_T$ jets in Section III.  We note that jet spectra
carry the signature of the hard-scattering process with a power index
$n$~$\sim$ 4 - 5 at central rapidity.  In Section IV, we study the UA1
and ATLAS data which extend from the low-$p_T$ region of a few GeV to the
high-$p_T$ region up to  150 GeV.  We find that the power index for jet
production is approximately 4 - 5, supporting the dominance of the
hard-scattering process over the extended $p_T$ region at central
rapidity.  In Section V, we seek additional evidences of the
hard-scattering process from two-particle correlation data.  In
Section VI, we study the effects of parton showering on the evolution
of the jet spectra to the hadron spectra.  In Section VII, we examine
the regularization and further approximation of the relativistic
hard-scattering integral to bring it to the form of the nonextensive
statistical mechanics.  In Section VIII, we analyze hadron spectra
using the nonextensive statistical mechanical distribution.  We
present our concluding summary and discussions in Section
IX.

\vspace*{-0.2cm}
\section{Approximate Hard-Scattering Integral
\label{sec2}}

We would like to review and summarize the results of the
hard-scattering integral obtained in our earlier works in
\cite{Won94,Won98,Won13,Won13Is,Won14EPJ} so that we can refine
previous analytical results.  We consider the collision of $A$ and $B$
in the center-of-mass frame at an energy $\sqrt{s}$ with the detected
particle $c$ coming out at $\eta\sim 0$ in the reaction $A+B \to c+X$
as a result of the relativistic hard-scattering of partons $a$ from
$A$ with parton $b$ from $B$.  Upon neglecting the intrinsic
transverse momentum and rest masses, the differential cross section in
the lowest-order parton-parton elastic collisions is given by
\begin{eqnarray}
\frac{E_cd^3\sigma( AB \to c X) }{dc^3}
&&=\sum_{ab} \int dx_a  dx_b
G_{a/A}(x_a) G_{b/B} (x_b)
\nonumber\\
&&\times
 \frac{E_cd^3\sigma( ab \to c X') }{dc^3},
\label{2}
\end{eqnarray}
where we use the notations in Ref.\ \cite{Won13} with $c$ the momentum
of the produced parton, $x_a$ and $x_b$ the forward light-cone
variables of colliding partons $a$ and $b$ in $A$ and $B$, respectively
and $d\sigma( ab\!\! \to\!\! c X') /dt$ the parton-parton invariant
cross section.

We are interested in the production of particle $c$ at $\theta_{\rm
  CM}=90^o$ for which analytical approximate results can be obtained.
We integrate over $dx_a$ in Eq.\ (\ref{2}) by using the delta-function
constraint in the parton-parton invariant cross section, and we
integrate over $dx_b$ by the saddle-point method to write
\begin{eqnarray}
[x_{a0}G_{a/A}(x_{a})][ x_{b0}G_{b/B}(x_{b})]
=e^{f(x_b)}.
\end{eqnarray}
We expand $f(x_b)$ about its minimum at $x_{b0}$.  We obtain
\begin{eqnarray}
\int d x_b e^{f(x_b)} g(x_b)&\sim& e^{f(x_{b0})} g(x_{b0})\sqrt{
\frac{2 \pi}{-\partial ^2 f (x_{b}) /\partial x_b^2|_{x_b=x_{b0}}} }.
\nonumber
\end{eqnarray}
For simplicity, we assume $G_{a/A}$ and $G_{b/B}$ to have the same
form.  At $\theta_c\sim 90^0$ in the CM system, the minimum value of
$f(x_b)$ is located at
\begin{eqnarray}
x_{b0}=x_{a0}=2x_c,~~~{\rm and}~~~x_c=\frac{c_T}{\sqrt{s}}.
\label{5}
\end{eqnarray}
\vspace*{-0.5cm}
We get
\vspace*{0.3cm}
\begin{eqnarray}
E_C \frac{d^3\sigma( AB\!\! \to\!\! c X) }{dc^3}
&&\sim
\sum_{ab}
B [x_{a0}G_{a/A}(x_{a0})][ x_{b0}G_{b/B}(x_{b0})]
\nonumber\\
&&\times \frac{d\sigma(ab\! \to\! cX')}{dt},
\end{eqnarray}
where
\begin{eqnarray}
B=\frac{1}{\pi (x_{b0}-c_T^2/x_c s)}
\sqrt{
\frac{2 \pi}{-\partial ^2 f (x_{b}) /\partial x_b^2|_{x_b=x_{b0}}} }.
\end{eqnarray}
For the case of $x_aG_{a/A}(x_a)=A_a(1-x_a)^{g_a}$, we find
\begin{eqnarray}
-\frac{\partial ^2 f (x_{b})}{\partial x_b^2 }\biggr |_{x_b=x_{b0}}=\frac{2g(1-x_c)}{x_c(1-x_{a0})(1-x_{b0})},
\end{eqnarray}
and we obtain\footnote{ In Eq.\ (23) of the earlier work of
  \cite{Won13}, there was a typographical error in the quantity
  $x_{b0}$ in the denominator under the square root sign, which should
  be $x_c$.  }
\begin{eqnarray}
E_C \frac{d^3\sigma( AB\!\! \to\!\! c X) }{dc^3}
&&\sim
\sum_{ab}{ A_aA_b}
\frac{(1-x_{a0})^{g_a+\frac{1}{2}}(1-x_{b0})^{g_b+\frac{1}{2}}}
{\sqrt{\pi g_a}\sqrt{ x_c(1-x_c)}}
\nonumber\\
&&\times
\frac{d\sigma(ab\! \to\! cX')}{dt}.
\label{rhs}
\end{eqnarray}
The above analytical result differs from the previous result of Eq.\ (9)  of Ref.\ \cite{Won13}, where the factor that appears in the above equation
\begin{eqnarray}
{(1-x_{a0})^{\frac{1}{2}}(1-x_{b0})^{\frac{1}{2}}}/\sqrt{(1-x_c)}
\end{eqnarray}
was approximated to be unity.  We wish to retain such a factor in order to obtain a more accurate determination of the power index, in cases where $c_T$ may be a substantial fraction of $\sqrt{s}$.

If the basic process is $gg \to gg$, the cross section at
$\theta_c\sim 90^{o}$ \cite{Gas90} is
\begin{eqnarray}
\frac{d\sigma(gg\to gg) }{dt}
&\sim&
\frac{9\pi \alpha_s^2}{16c_T^4}
\left [\frac{3}{2} \right ]^3.
\label{36}
\end{eqnarray}
If the basic process is $qq' \to qq'$, the cross section
at $\theta_c\sim 90^{o}$ \cite{Gas90} is
\begin{eqnarray}
\frac{d\sigma(qq' \to qq')}{dt}
&=&
\frac{4 \pi \alpha_s^2}{9c_T^4}
\frac{5}{16}.
\label{38}
\end{eqnarray}
If the basic process is $gq\to gq'$, the cross section
at $\theta_c\sim 90^{o}$ \cite{Gas90} is
\begin{eqnarray}
\frac{d\sigma(gq \to gq)}{dt}
&=&
\frac{5 \pi \alpha_s^2}{4c_T^4}
\frac{11}{36}.
\label{39}
\end{eqnarray}
In all cases, the differential cross section varies as
$d\sigma(ab\!\!\to\!\! cX')/dt \sim \alpha_s^2/(c_T^2)^2$.

\section{The Power Index in Jet Production at high $p_T$
\label{sec4}}

Our earlier investigation on the effects of multiple collisions
indicates that without a centrality selection in minimum-biased
events, the differential cross section for the production of partons
at high-$p_T$ will be dominated by the contribution from a single
parton-parton scattering that behaves as $1/c_T^4$
\cite{Kas87,Cal90,Gyu01,Cor10,Won13}. It suffices to consider only the
results of the single parton-parton collision as given in
Eq.\ (\ref{rhs}) which can be compared directly with the transverse
differential cross sections for hadron jet and isolated photon
production.

From the results in the parton-parton cross sections in
Eqs.\ (\ref{36},\ref{38},\ref{39}), the approximate analytical formula
for hard-scattering invariant cross section $\sigma_{\rm inv}$, for
$A+B \to c+X$ at $\eta\sim 0$, is
\begin{eqnarray}
 &&\hspace*{-0.6cm}E_c \frac{d^3\sigma( AB\!\! \to\!\! c X) }{dc^3}\biggr |_{\eta\sim 0}
\nonumber\\
&&
\propto
\frac{\alpha_s^2 (1-x_{a0}(c_T))^{g_a+\frac{1}{2}}(1-x_{b0}(c_T))^{g_b+\frac{1}{2}}}
{c_T^{4}\sqrt{c_T/\sqrt{s}}  \sqrt{1-x_c(c_T)}}.
\label{19}
\end{eqnarray}
We analyze the $c_{{}_T}$ spectra by using a running
coupling constant
\begin{eqnarray}
\alpha_s(Q(c_T)) = \frac{12\pi}{27 \ln(C + Q^2/\Lambda_{\rm QCD}^2)},
\label{run}
\end{eqnarray}
where $\Lambda_{\rm QCD}$ is chosen to be 0.25 GeV to give
$\alpha_s(M_Z^2)=0.1184$ \cite{Ber12}, and the constant $C$ is chosen
to be 10, both to give $\alpha_s(Q$$\sim$$\Lambda_{\rm QCD})$ $\sim$
0.6 in hadron spectroscopy studies \cite{Won01} and to regularize the
coupling constant for small values of $Q(c_{{}_T})$.  We identify $Q$
as $c_{{}_T}$ and search for $n$ by writing the invariant cross
section for jet production as
\begin{eqnarray}
& & \hspace{-0.6cm}\sigma_{\rm inv}\equiv E_c \frac{d^3\sigma( AB\!\! \to\!\! c X) }{dc^3}\biggr |_{\eta\sim 0}~~~~
\nonumber\\
&&
\hspace{-0.6cm}=\! \frac{A\alpha_s^2(Q^2(c_T)) (1-x_{a0}(c_T))^{g_a+\frac{1}{2}}(1-x_{b0}(c_T))^{g_b+\frac{1}{2}}}
{c_T^{n} \sqrt{1-x_c(c_T)}},
\label{22}
\end{eqnarray}
where  the power index $n$ for perturbative QCD has the value of 4.5.

\begin{figure} [h]
\hspace*{-0.4cm}
\includegraphics[scale=0.33]{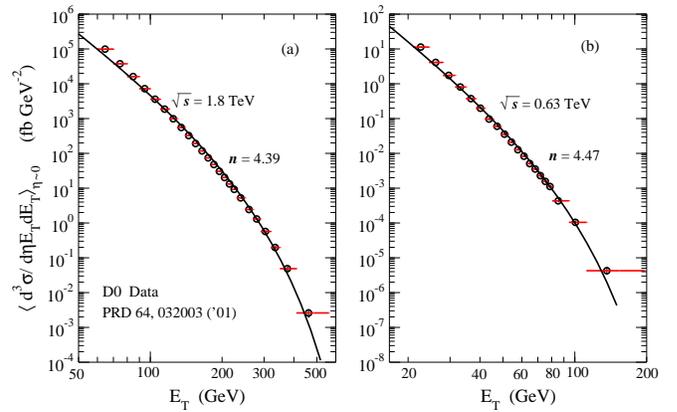}
\caption{(Color online) Comparison of the relativistic hard-scattering
  model results for jet production, Eq.\ (\ref{22}) (solid curves),
  with experimental $d\sigma/d\eta E_T dE_T|_{\eta\sim 0}$ data from the D0
  Collaboration \cite{Abb01}, for hadron jet production within
  $|\eta|$$<$0.5, in $\bar p p$ collision at (a) $\sqrt{s}$=1.80 TeV,
  and (b) $\sqrt{s}$=0.63 TeV. }
\label{fig1}
\end{figure}

We identify the parton $c$ with the produced jet and we define the jet
transverse rapidity $y_T$ as the logarithm of $c_T/\sqrt{s}$,
\begin{eqnarray}
y_T=\ln \left ( \frac{c_T}{\sqrt{s}}\right ),~~~ e^{y_T} =\frac{ c_T}{\sqrt{s}},
\end{eqnarray}
then the results in Eq.\ (\ref{22}) gives
\begin{eqnarray}
\hspace*{-0.5cm}\frac{\partial \ln \sigma_{\rm inv} }{dy_T} \!=\! \frac{\partial \alpha_s}{dy_T}\! -\! \frac{2(g_a+g_b+1)e^{y_T} }{1-2 e^{y_T} }  \!-\! n + \frac{e^{y_T} }{2(1-e^{y_T} )},
\end{eqnarray}
and
\begin{eqnarray}
\hspace*{-0.5cm}\frac{\partial^2  \ln \sigma_{\rm inv}}{dy_T^2}
\!=\! \frac{\partial^2 \alpha_s}{dy_T^2} \!-\!  \frac{2(g_a+g_b+1)e^{y_T}}{(1-2e^{y_T})^2}
 \!+\! \frac{e^{y_T}}{2(1-e^{y_T})^2}.
\end{eqnarray}
Therefore in the (ln $\sigma_{\rm inv}$)-(ln $E_T(c_T$)) plot in
Fig.\ \ref{fig1}, the slope $({\partial \ln \sigma_{\rm inv} }/{dy_T})$ at
small values of $E_T$ gives approximately the power index $n$ and the
second derivative of $\ln \sigma_{\rm inv}$ with respect to $\ln E_T$
at large values of $E_T$ gives approximately the power index
$g_a$+$g_b$ of the structure function. The exponential index $g_a=g_b$
for the structure function of a gluon varies from 6 to 10 in different
structure functions \cite{Duk84,Che03}.  We shall take $g_a=6$ from
\cite{Duk84}.

With our refinement of the hard-scattering integral in Eq.\ (9), our analytical invariant cross section of Eq.\ (\ref{22}) differs from that in our earlier work in \cite{Won13} in the presence of an extra energy- and $p_T$-dependent factor of Eq.\ (10) and a slightly different running  coupling constant.  
We shall re-examine  the power indices with Eq.\ (\ref{22}).  We wish  to 
obtain a more accurate determination of the power indices, in cases   where $c_T$ may be a substantial fraction of the collision energy $\sqrt{s}$.
We also wish to use    Eq.\ (\ref{22}) to calibrate the signature of the power indices for jets at high $p_T$, where  jets can be better isolated, to test 
in the next section 
the power indices for jets  extending to the  lower $p_T$ region,
where jets are more numerous and harder to isolate.

Using Eq.\ (\ref{22}), we find that the  $d\sigma/d\eta E_T dE_T|_{\eta\sim 0}$ data from the D0  Collaboration \cite{Abb01}
for hadron jet production within $|\eta|$$<$0.5 can be fitted with $n$=4.39 for $\bar p p$ collisions at
$\sqrt{s}$=1.8 TeV, and with $n$=4.47 for $\bar p p$ collisions at
$\sqrt{s}$=0.630 TeV, as shown in Fig.\ {\ref{fig1}}.

In another comparison with the ALICE data for jet production in $pp$
collisions at $\sqrt{s}=2.76$ TeV at the LHC within $|\eta|<0.5$
\cite{Alice13} in Fig.\ 2, the power index is $n$=4.78 for $R=0.2$,
and is $n$=4.98 for $R=0.4$ (Table I).  In Fig.\ 3, the power index is
$n$=5.39, for CMS jet differential cross section in $pp$ collisions at
$\sqrt{s}=7$ TeV at the LHC within $|\eta|<0.5$ and $R=0.5$
\cite{cms11}.  
\begin{figure} [h]
\hspace*{-0.6cm}
\includegraphics[scale=0.37]{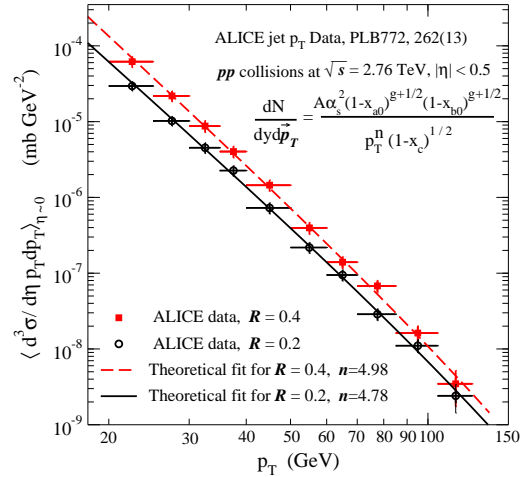}
\caption{ (Color online) Comparison of the relativistic hard-scattering
  model results for jet production, Eq.\ (\ref{22}) (solid curves),
  with experimental $d\sigma/d\eta E_T dE_T|_{\eta\sim 0}$ data from the ALICE
  Collaboration \cite{Alice13}, for jet production within
  $|\eta|$$<$0.5, in $p p$ collision at 2.76 TeV for $R$=0.4, and
  $R$=0.2. }
\label{fig2}
\end{figure}

The power indices extracted from the  hadron jet spectra in D0 \cite{ Abb01}, \cite{Alice13}, and CMS \cite{cms11} are  listed in
Table I.  
The extracted D0 power indices  are smaller than  those extracted
previously in 
\cite{Won13} by 0.2 units, as the highest transverse momenta are substantial fractions of the collision energy.  In the other cases, the change of the power indices from our earlier work in \cite{Won13} are small
as their highest  transverse momenta are substantially smaller than the collision energies. 

\begin{figure} [h]
\hspace*{-0.6cm}
\includegraphics[scale=0.35]{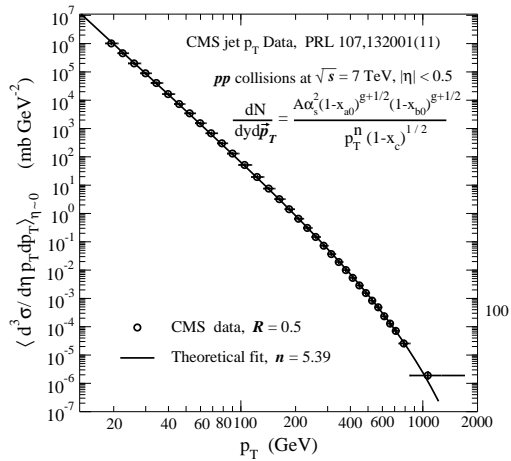}
\caption{Comparison of the relativistic hard-scattering
  model results for jet production, Eq.\ (\ref{22}) (solid curves),
  with experimental $d\sigma/d\eta E_T dE_T|_{\eta\sim 0}$ data from the CMS
  Collaboration \cite{cms11}, for jet production within
  $|\eta|$$<$0.5, in $p p$ collision at 7 TeV. }
\end{figure}

With the jet spectra  at high $p_T$ from the D0, ALICE and CMS 
Collaborations, we find that the signature for jet production 
is a power index in the range of  4.5  to 5.4, with  a small variation that depend on  $\sqrt{s}$  and $R$
as shown in Table I. 
While these power indices are close to the lowest-order pQCD prediction of 4.5, 
there appears to be a 
consistent 
tendency for $n$ to increase slightly as $\sqrt{s}$  and  $R$ increases.  Such an increase may arise from higher-order pQCD effects.
 We can envisage the physical picture that as the jet evolves by parton showering,
the number of generations of parton branching will increase with a greater collision energy $\sqrt{s}$  or  a greater opening angle  $R$.
 A greater $\sqrt{s}$  or a  larger  $R$ value
corresponds to a later stage of  the evolution of the parton showering and they will lead naturally to a slightly greater value of the power index $n$.

\begin{table}[h]
\caption { The power index $n$ for the jet spectra in
$\bar p p$ and $pp$  collisions.  Here, $R$ is the jet cone angular radius used in the jet search algorithm. }
\vspace*{0.2cm}
\begin{tabular}{|c|c|c|c|c|c|}
\cline{2-6}
  \multicolumn{1}{c|}{}        &  $\sqrt{s}$ & $p_T$ Region & $R$  &  $\eta$   & $n$
  \\
\cline{1-6}
D0\cite{Abb01}&$\bar p p$ 1.80TeV\!\!&\!\!64$<$$p_T$$< $461GeV\!\!& 0.7  &     $|\eta|$$<$0.7& 4.39
 \\ \hline
D0\cite{Abb01}&$\bar p p$ 0.63TeV\!\!&\!\!22$<$$p_T$$<$136GeV\!\!&  0.7  &  $|\eta|$$<$0.7& 4.47\\ \hline
ALICE\cite{Alice13}\!\!& $p p$ 2.76TeV\!\!&\!\!22$<$$p_T$$< $115GeV\!\!&  0.2  & $|\eta|$$<$0.5& 4.78
 \\ \hline
ALICE\cite{Alice13}\!\!&$p p$ 2.76TeV\!\!&\!\!22$<$$p_T$$< $115GeV\!\!&  0.4  &  $|\eta|$$<$0.5  & 4.98\\ \hline
CMS \cite{cms11}&$p p$  7TeV\!\!&\!19$<$$p_T$$< $1064GeV\!\!&  0.5  &  $|\eta|$$<$0.5& 5.39\\ \hline
\end{tabular}
\end{table}

The signature of the power indices for the production of jets at 
high $p_T$ can be used to identify the nature of the  jet production process at low $p_T$.   If the power indices in the production in the  lower-$p_T$ region are similar, then the jets in the lower-$p_T$ region and the jets in the high-$p_T$ region have the same spectral shape  and
can be considered to originate from the same production mechanism, extending the dominance of the  relativistic hard-scattering process to the lower-$p_T$ region.

\section{Jet Production in an Extended  Region from low to high $p_T$ }

The analysis in the last section was carried out for jets with a
transverse momentum greater than 19 GeV.  It is of interest to find
out whether the perturbative QCD power index remains a useful concept
when we include also the production of jet-like energy clusters
(mini-jets) at lower transverse momenta. In order to apply the
power-law (\ref{22}) to the whole range of $c_T$, we need to
regularize it by the replacement\footnote{ So far, the only rationale
  behind this is that, in the QCD approach, large $c_T$ partons probe
  small distances (with small cross sections).  With diminishing of
  $c_T$ , these distances become larger (and cross sections are
  increasing) and, eventually, they start to be of the order of the
  nucleon size (actually it happens around $c_T \simeq c_{T0}\sim
  1/r_{\rm nucleon}$ or $m_T \simeq m_{T0}$ ). At that point the cross
  section should stop rising, i.e., it should not depend anymore on
  the further decreasing of transverse momentum $c_T$. The scale
  parameter $m_{T0}$ can then be identified with $m_{T0}$
  here. Similar idea was employed when proposing Eq. (\ref{run}).},
\begin{equation}
\frac{1}{c_T^2}  \to \frac{1}{ 1+{m_T^2}/{m_{T0}^2}}.
\label{18}
\end{equation}
or alternatively as
\begin{equation}
\frac{1}{c_T}  \to \frac{1}{ 1+{m_T}/{m_{T0}}}.
\label{19a}
\end{equation}
The quantity $m_{T0}$ measures the average transverse mass of the
detected jet in the hard-scattering process.  Upon choosing the
regularization (\ref{18}), the differential cross section $d^3\sigma(
AB \to p X) /dy d{\bb p}_T$ in (\ref{22}) is then regularized as
\begin{eqnarray}
\!\!\!\!\!\!\!\!\!\!\!\!\!\!\!  \frac{d^3\sigma( AB \to p X) }{dy
d{\bb
p}_T} \biggr | _{ y \sim 0} && \nonumber\\
&& \hspace{-3.0cm} \propto \frac{\alpha_s^2(Q^2(
c_T))(1\!-\!x_{a0}( c_T))^{g_a+1/2}(1\!-\!x_{b0}(
c_T))^{g_b+1/2} } {[1+m_{T}^2(c_T)/m_{T0}^2]^{n/2}\sqrt{1-x_c( c_T)}}
. \label{30}
\end{eqnarray}
We fit the inclusive UA1 jet cross sections at $\eta\sim 0$
\cite{UA188} as a function of the jet $p_T$ for $p\bar p$ collisions
with the above equation, and we find that $n$=4.47 and $m_{T0}$=0.267
GeV for $p\bar p$ collisions at $\sqrt{s}$=564 GeV, and $n$=4.73 and
$m_{T0}$=0.363 GeV for $p\bar p$ collisions at $\sqrt{s}$=630 GeV.

\begin{figure} [h]
\hspace*{-0.5cm}
\includegraphics[scale=0.45]{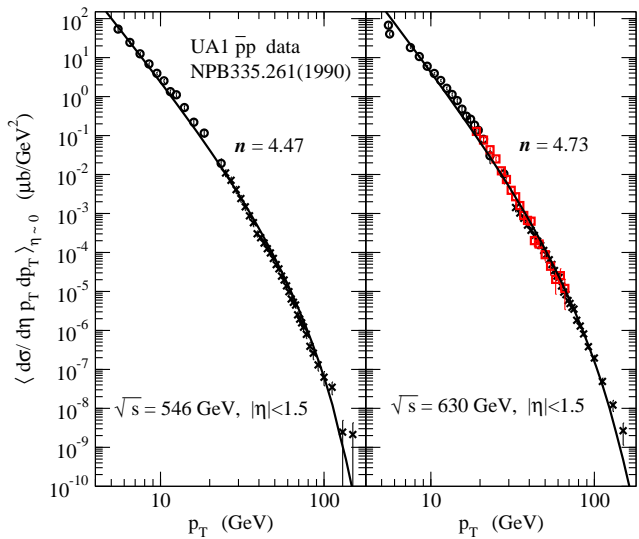}
\caption{(Color online) Comparison of the relativistic hard-scattering
  model results for jet production, Eq.\ (\ref{30}) (solid curves),
  with experimental $d\sigma/d\eta\, p_T dp_T$ data from the UA1
  Collaboration \cite{UA188}, for jet
 production within
  $|\eta|$$<$1.5, in $\bar p p$ collision at (a) $\sqrt{s}$=0.546 TeV,
  and (b) $\sqrt{s}$=0.63 TeV. }
\end{figure}

\begin{figure} [h]
\hspace*{-0.5cm}
\includegraphics[scale=0.45]{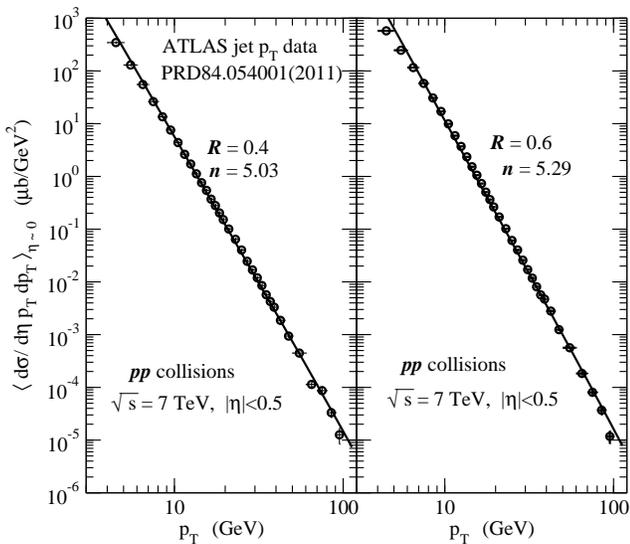}
\caption{Comparison of the relativistic hard-scattering
  model results for jet production, Eq.\ (\ref{30}) (solid curves),
  with experimental $d\sigma/d\eta\, p_T dp_T|_{\eta\sim 0}$ data from the  ATLAS
  Collaboration \cite{Aad11}, for jet
 production within
  $|\eta|$$<$0.5, in $ p p$ collision at $\sqrt{s}$=7 TeV, with (a)  $R$=0.4
  and (b) $R$=0.6. }
\label{ATL}
\end{figure}

The ATLAS $p_T$ spectra for $pp$ collisions at 7 TeV also extend to the region of a few 
GeV.  It is of interest to find out  what are the power indices for these collisions.  We show in Fig.\ {\ref{ATL}} the comparison of the results of Eq.\ (\ref{30}) with the ATLAS data at $\eta\sim$ 0  \cite{Aad11}.  We find that
$n$=5.03 for $R=0.4$ and $n=$ 5.29  for $R$=0.6.  Because the data start with $p_T$ of a few GeV, the fits and the extracted value of $n$ are insensitive to the variation of the  $m_{T0}$ values so that  there is an ambiguity in the product of the normalization and  $m_{T0}$ in the analysis.
The fits in Fig.\  (\ref{ATL}) have been obtained with $m_{T0}=$ 1 GeV.   The value of $n$ is  is related to the slope of the curves in Fig.\ \ref{ATL}.

\begin{table}[h]
\caption { The power index $n$ extracted from jet production in
$\bar p p$ and $pp$  collisions in the extended $p_T$ region from a few GeV to the 
high-$p_T$ region. }
\vspace*{0.2cm}
\begin{tabular}{|c|c|c|c|c|c|}
\cline{2-6}
  \multicolumn{1}{c|}{}        &  $\sqrt{s}$ & $p_T$ Region & $R$  &  $\eta$   & $n$
  \\
\cline{1-6}
UA1\cite{UA188}&$\bar p p$ 0.564TeV\!\!&\!5.5$<$$p_T$$< $150GeV\!\!&0.75&$|\eta|$$<$1.5& 4.47\\ \hline
UA1\cite{UA188} &$\bar p p$ 0.63TeV\!\!&\!5.5$<$$p_T$$< $150GeV\!\!&0.75&$|\eta$$<$1.5   & 4.73\\ \hline
\!ATLAS\cite{Aad11}\!&$p p$ 7TeV\!\!&\!\!4.5$<$$p_T$$< $95GeV\!\!&  0.4  &$|\eta|$$<$0.5& 5.03\\ \hline
\!ATLAS\cite{Aad11}\!&$p p$ 7TeV\!\!&\!\!4.5$<$$p_T$$< $95GeV\!\!&  0.6  &  $|\eta|$$<$0.5& 5.29\\ \hline
\end{tabular}
\end{table}

We list in Table II the power indices extracted from UA1 and ATLAS for the extended $p_T$ region from  a few GeV to the high-$p_T$ region.  It should be mentioned that the importance of the production of jets with $p_T$ of a few GeV (minijets) has already been  emphasized in the earlier work of \cite{Wan91}.

By comparing the power indices obtained in Table I for D0, ALICE, and CMS for jets at high $p_T$ with those for UA1 and ATLAS  for jets in  the lower-$p_T$ region in Table II, we note that these power indices are very similar.   The corresponding  power index values  are nearly the same, and the changes of the power index with respect to $\sqrt{s}$ and $R$ are nearly the same.  
They  can be considered to originate from  the same relativistic  hard-scattering mechanism, indicating the dominance of the hard-scattering process over the extended $p_T$ region
from a few GeV to about 100 GeV.

\section{Additional evidences of jet production from two-particle correlation data}

In addition to the spectral shape, we seek additional evidences of jet production in the low-$p_T$ region
from experimental two-particle correlations measurements, which
consist of correlations on the near-side and the away-side.  We shall
first examine near-side correlations .

The experimental distribution of near-side particles 
associated
with a trigger particle of momentum $p_T^{\rm trig}$ 
 in $pp$ collisions 
can be described well by \cite{Won08,Won09}
\begin{eqnarray}
\label{jetfun}
\frac { dN_{\rm jet}^{pp}} {p_T dp_T\, d\Delta \eta\, d\Delta \phi}
=&& N_{\rm jet}
\frac{\exp\{(m-\sqrt{m^2+p_T^2})/T_{\rm jet}\}} {T_{\rm jet}(m+T_{\rm jet})}
\nonumber\\
\times
&&\frac{1}{2\pi R^2}
e^{- {[(\Delta \phi)^2+(\Delta \eta)^2]}/{2R^2} },
\label{21}
\end{eqnarray}
where by assumption of hadron-parton duality $m$ can be taken as the
pion mass $m_\pi$, $N_{\rm jet}$ is the total number of near-side
(charged) associated particles in a $pp$ collision, and $T_{\rm jet}$
is the jet inverse slope (``temperature'') parameter of the ``$pp$ jet
component''.  We find that the parameters $N_{\rm jet}$ and $T_{\rm
  jet}$ vary linearly with $p_T^{\rm trig}$ of the trigger particle
which we describe as
\begin{eqnarray}
\label{Njetv}
N_{\rm jet} = N_{{\rm jet}0} + d_N ~ p_T^{\rm trig},
\label{22a}
\end{eqnarray}
\begin{eqnarray}
\label{Tjetv}
T_{\rm jet}= T_{{\rm jet}0} + d_T ~ p_T^{\rm trig}.
\end{eqnarray}
We also find that the width parameter $R$ depends slightly on $p_T$
which we can parametrize as
\begin{eqnarray}
\label{ma}
R=R_0 \frac{m_a}{\sqrt{m_a^2+p_T^2}}.
\end{eqnarray}
\begin{figure} [h]
\hspace*{-0.5cm}
\includegraphics[angle=0,scale=0.43]{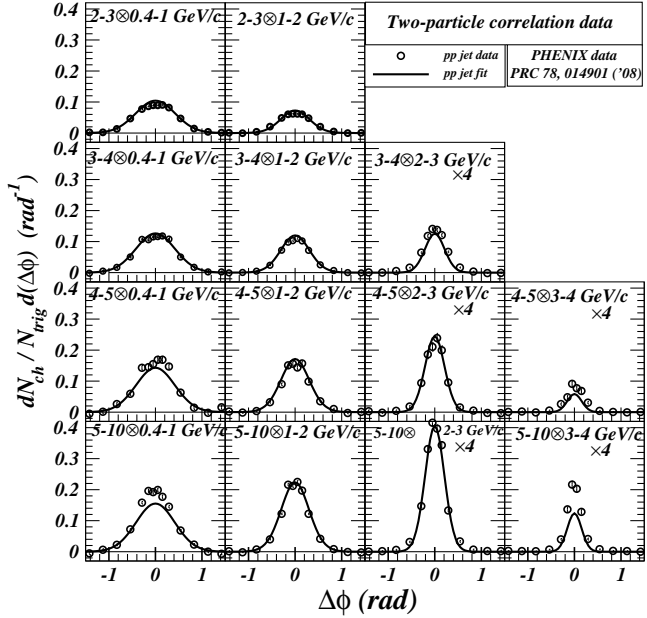}
\vspace*{0.0cm}
\caption{ PHENIX azimuthal angular distribution of
  associated particles per trigger in different $p_t^{\rm trig}\otimes
  p_t^{\rm assoc}$ combinations.  The open circles are the associated
  particle yields per trigger, $dN_{\rm ch}/N_{\rm trig} d\Delta \phi$,
  in $pp$ collisions \cite{PHEN08}.  The solid curves are the
  theoretical associated particle yields per trigger calculated with
  Eq.\ (\ref{21}) . }

\end{figure}
Using this set of parameters and Eq.\ (\ref{21}), we fit the $pp$
associated particle data obtained in PHENIX measurements for $pp$
collisions at $\sqrt{s_{NN}}=200$ GeV.  The values of the parameters
are given in Table III.  As extracted from Fig.\ 1 of \cite{Won09}, the
theoretical results of $dN_{\rm ch}^{pp}/N_{\rm trig} d\Delta\phi$
from Eq.\ (\ref{21}) are given as solid curves in Fig. 6 and the
corresponding experimental data are represented by open circles.  As
one observes in Fig.\ 6, although the fit is not perfect, the set of
parameters in Table III adequately describe the set of $pp$ associated
particle data for $2< p_T^{\rm trig} < 10$ GeV and for $0.4 < p_T^{\rm
  assoc}<4$ GeV.  As indicated in Table III, the parameters of
Eqs.\ (\ref{Njetv}) and (\ref{Tjetv}) are $N_{{\rm jet}0}=0.15$,
$d_N=0.1/$GeV, $T_{{\rm jet}0}=0.19$ GeV, and $ d_T=0.06$.
It is interesting to note that the cone angle $R_0$ for jets in the lower-$p_T$ region is of the same order as those in the high-$p_T$ region.

\begin{table}[h]
\caption { Jet component parameters in Eq.\ (\ref{jetfun}) obtained for the description of
experimental near-side 
associated particles with different $p_t^{\rm trig}$ triggers in STAR \cite{STAR05} and PHENIX \cite{PHEN08} Collaborations, in $pp$
collisions at $\sqrt{s_{NN}}$=200 GeV.}
\vspace*{0.3cm}
\begin{tabular}{|c|c|c|c|c|c|}
\cline{2-6}
 \multicolumn{1}{c|}{}   &  STAR &  \multicolumn{4}{c|} {PHENIX} \\ \hline
 $p_T^{\rm trig}$ &  4-6GeV & 2-3GeV & 3-4GeV & 4-5GeV  & 5-10GeV  \\ \hline
 $N_{\rm jet}$ & 0.75  &
\multicolumn{4}{c|} {0.15+0.10 $\langle p_T^{trig} \rangle/{\rm GeV}$ }
\\ \cline{1-6}
 $T_{\rm jet}$ &{ 0.55{\rm GeV}}  &
\multicolumn{4}{c|} {0.19 GeV+0.06 $\langle p_T^{trig} \rangle $}
\\ \cline{1-6}
 $R_0$ & \multicolumn {5} {c|} {0.50}
\\ \cline{1-6}
$m_a$ & \multicolumn{5} {c|} {1.1 GeV} \\ \hline
\end{tabular}
\end{table}

\begin{figure} [h]
\hspace*{-0.5cm}
\includegraphics[angle=0,scale=0.70]{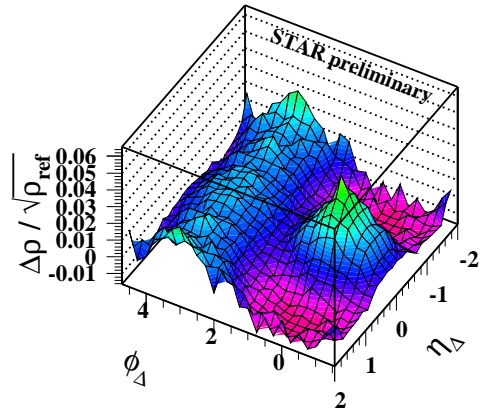}
\vspace*{-10.0cm}
\caption{(Color online) Minimum-$p_T$-biased two-particle angular
  correlation, without a $p_T$ trigger selection, for charged hadrons produced
  in $pp$ collisions at $\sqrt{s}$=200 GeV (from Fig.\ 1 of
  Ref.\ \cite{Por05} of the STAR Collaboration). Here,  $\phi_\Delta$ is $\Delta \phi$, the difference of the azimuthal  angles of two detected charged hadrons, and $\eta_\Delta$ is $\Delta \eta$, the difference of their pseudorapidities. }
\label{fig6}
\end{figure}

The presence of a well-defined cone of particles associated
with a $p_T>$ 2-3 GeV triggers in Fig.\ 6 on the near-side and the non-vanishing
extrapolation of the jet yield $N_{\rm jet}$ to the case of a
low-$p_T$ trigger in Eq.\ (\ref{22a}) provide an additional evidence
of jet production in the $p_{T ~\rm trigger} >$ 2 GeV region
in high energy $pp$ collisions. Furthermore, 
even in minimum-$p_T$-biased events without a high-$p_T$ trigger,
a similar cone of
associated correlated particles at $(\Delta \phi, \Delta \eta) \sim $
0 are present in two-particle correlation data,  as shown in
Fig.\ \ref{fig6} \cite{STAR06,STAR06twopar,Por05,Tra11},
indicating the production of jet-like structure on the near-side for
low-$p_T$ particles.

 In addition to the particles associated with trigger particle on the
 near-side, there are particles associated with the trigger particle
 on the back-to-back, away-side at $\Delta \phi$$\sim$$\pi$, in the
 form of a ridge along the $\Delta\eta$ direction, both with
 high-$p_T$  \cite{STAR05,STAR06,Put07} and low-$p_T$
 triggers \cite{STAR06,STAR06twopar,Por05,Tra11}, for $pp$
 collisions at $\sqrt{s}$= 200 GeV.  
Here,  the importance of the production  the low-$p_T$ jet (minijets) in the low-$p_T$ region has already been pointed out previously in the work of \cite{ STAR06twopar,Por05,Tra11,Ray11}. 
In Fig.\ 7, (taken from the STAR data in Fig. 1 of \cite{Por05}), we show the
 two-particle correlation data in a minimum-$p_T$-biased measurements
 which corresponds
 to the case with a low-$p_T$ trigger. The two-particle correlation
 data in Fig.\ \ref{fig6} indicate the presence of (i) a near-side
 particle cluster at $(\Delta \phi, \Delta \eta )\sim 0$ (a mini-jet)
 and (ii) an away-side ridge of associated particles at $\Delta
 \phi\sim \pi$.    The $\Delta\phi\sim \pi$ (back-to-back)
 correlation in the shape
 of a ridge indicates that the two particles are parts of the partons from the two
 nucleons and they carry fractions of the longitudinal momenta of
 their parents, leading to the ridge of~$\Delta \eta$ at
 $\Delta\phi\sim \pi$.  These two group of particles at $\Delta \phi
 \sim 0$ and $\Delta \phi\sim \pi$ can be interpreted as arising from the pair of scattered
 partons in a relativistic hard scattering.

The dominance of the hard scattering in the spectrum does not imply
the absence of soft processes.  It only stipulates that the soft
process contribution is much smaller in comparison.  In the lowest
$p_T$ region, one expects contributions from soft nonperturbative QCD
physics that may involve the parton wave functions in a flux tube
\cite{Gat92}, the thermodynamics and the recombination of partons
\cite{H,HR,Kra09,Hwa03}, or the fragmentation of a QCD string
\cite{And83,Sch51,Wan88}. However, as the contributions from the
hard-scattering processes increase with increasing collision
energies, the fraction of the contributions from soft processes becomes smaller in
comparison with the contributions from the hard-scattering processes,
as pointed out earlier in \cite{UA188,Wan91}.  As a consequence, the
contributions from the hard-scattering process can dominate the particle production process in  high-energy $pp$ and $p\bar p$ collisions.

\section{Effects of Parton Showering  on Transverse Differential Cross Section}

The last sections show the possible dominance of jet production at
central rapidity in high-energy $pp$ and $p\bar p$ collisions over essentially the whole 
 $p_T$ region. We would like to find out how the jets evolve
to become hadrons and how the hadron spectra manifest themselves.

In addition to the jet transverse spectra, experimental measurements
also yield the hadron spectra without the reconstruction of jets. The
hadron transverse spectra give a slightly greater power index, $n_{\rm
  hadron}$$\sim$6-10
\cite{Ang78,TR13,Arl10,Won12,Won13,Won13Is,Won14EPJ}. Previously, we
outline how the increase in the power index $n$ from jet production to
hadron production may arise from the subsequent parton showering that
turns jets into hadrons \cite{Won13}.  We would like to describe here
the evolution in more details. To distinguish between jets and its
shower products, we shall use the symbol $c$ to label a parent parton
jet and its momentum and the symbol $p$ to label a shower product
hadron and its momentum.

The evolution of the parton jet into hadrons by parton showering has
been described well by many models \cite {Sjo05}. There are three
different parton showering schemes: the PYTHIA \cite{PYTHIA}, the
HERWIG \cite{HERWIG}, and the ARIADNE \cite{ARIADNE}.  The general
picture is that the initial parton is characterized by a momentum and
a virtuality which measures the degree of the parton to be
off-the-mass-shell.  The parton is subject to initial-state and
final-state radiations. After the hard scattering process, the parton
possesses a high degree of virtuality $Q^{(0)}$, which can be
identified with the magnitude of the parton transverse momentum
$c_T$. The final-state radiation splits the parton into binary quanta
as described by the following splitting DGLAP kernels \cite{DGLAP},
\begin{eqnarray}
P_{q\to qg}&=&\frac{1}{3} \frac{1+z^2}{1-z},\\
P_{g\to gg}&=&3 \frac{[1-z(1-z)]^2}{z(1-z)},\\
P_{g\to q\bar q}&=&\frac{n_f}{2} [ z^2 +(1-z)^2] ,
\end{eqnarray}
where $z$ is the momentum fraction of one of the showered partons,
and there is symmetry between $z$ and $1-z$  for symmetrical
products in the second and third processes. After  the showering
splitting processes, there is always a leading parton with
\begin{eqnarray}
z_{\rm leading} \gg z_{\rm non-leading}.
\end{eqnarray}

For the study of the $p_T$ hadron spectra as a result of the parton
showering, it suffices to focus attention on the leading parton after
each showering splitting because of the rapid fall-off of the
transverse momentum distribution as a function of increasing $c_T$.
As a consequence, we can envisages the approximation conservation of
the leading parton as the parton showering proceeds and as its
momentum is degraded in each showering branching by the fraction
$\langle z \rangle$=$\langle z_{\rm leading} \rangle$.  In the present
study of high-$p_T$ particles in the central rapidity region, the
parton $c$ is predominantly along the transverse direction, and the
showering of the produced hadrons will also be along the transverse
direction.

A jet parton $c$ which evolves by parton showering will go through many
generations of showering.  If we label the (average) momentum of the
$i$-th generation parton by $c_T^{(i)}$, the showering can be
represented as $c_T \to c_T^{(1)} \to c_T^{(2)} \to c_T^{(3)} \to
...\to c_T^{(\lambda)}$=$p_T$.  Each branching will kinematically
degrade the momentum of the showering parton by a momentum fraction,
$\langle z \rangle $=$ c_T^{(i+1)}/ c_T^{(i)}$.  At the end of the
terminating $\lambda$-th generation of showering, the jet hadronizes
and the $p_T$ of a produced hadron is related to the $c_T$ of the
parent parton jet by
\begin{eqnarray}
\frac{p_T}{c_T} \equiv \frac{c_T^{(\lambda)}}{c_T} =\langle z
\rangle^\lambda .
\label{eq52}
\end{eqnarray}
It is easy to prove that if the generation number $\lambda$ and the
fragmentation fraction $z$ are independent of the jet $c_T$, then the
power law and the power index for the $p_T$ distribution are unchanged
\cite{Won13}.

We note however that in addition to the kinematic decrease of $c_T$ as
described by (\ref{eq52}), the showering generation number $\lambda$
is governed by an additional criterion on the virtuality.  From the
different parton showering schemes in the PYTHIA \cite{PYTHIA}, the
HERWIG \cite{HERWIG}, and the ARIADNE \cite{ARIADNE}, we can extract the
general picture that the initial parton with a large initial
virtuality $Q^{(0)}$ decreases its virtuality by showering until a
limit of $Q^{\rm cutoff}$ is reached.  The downgrading of the
virtuality will proceed as $Q^{\rm jet}$=$Q^{(0)} \to Q^{(1)} \to
Q^{(2)} \to Q^{(3)} \to ... \to Q^{(\lambda)}$=$Q^{\rm cutoff}$, with
\begin{eqnarray}
\langle \xi \rangle =\frac { Q^{(i+1)}}{Q^{(i)}} ~~{\rm and~~}
 \frac{Q^{(\lambda)}}{Q^{\rm jet}}=\langle \xi \rangle^{\lambda}.
\label{31}
\end{eqnarray}

The measure of virtuality has been defined in many different ways in
different parton showering schemes.  We can follow PYTHIA \cite{Sjo05}
as an example.  We consider a parton branching of $a \to bc$.  The
transverse momentum along the jet $a$ direction is
\begin{eqnarray}
b_T^2 &&=z(1-z)a^2- (1-z)b^2 -z c^2.
\end{eqnarray}
If $a^2=[Q^{(i)}]^2$=virtuality before parton branching, and $b^2=c^2=0$, as is assumed by PYTHIA, then
\begin{eqnarray}
b_T^2 &&=[Q^{(i+1)}]^2=z(1-z)a^2=z(1-z)[Q^{(i)}]^2.
\end{eqnarray}
So, if we identify the transverse momentum $b_T^2$ along the jet axis
as the square of virtuality $[Q^{(i+1)}]^2$ after parton branching,
the quantity $z(1-z)$ measures the degradation of the square of the
virtuality in each QCD branching process,
\begin{eqnarray}
\frac{[Q^{(i+1)}]^2}{[Q^{(i)}]^2}=z(1-z).
\end{eqnarray}
Thus,  the virtuality  fraction  of Eq.\ (\ref{31}) is related to
$\langle z(1-z) \rangle $  by
\begin{eqnarray}
\langle \xi \rangle = \sqrt{\langle z(1-z) \rangle }.
\end{eqnarray}
As $z$ is less than 1, $\langle \xi \rangle < \langle z \rangle$ which
implies that on the average the virtuality fraction $\langle \xi
\rangle$ in a parton branching is smaller than the momentum fraction
$\langle z \rangle$.  As a consequence, the virtuality of the leading
parton is degraded faster than its momentum as the showering process
proceeds so that when the virtuality reaches the cutoff limit, the
parton still retains a significant fraction of the initial jet
momentum.

The process of parton showering will be terminated when the virtuality
$Q^{(\lambda)}$ reaches the cutoff value $Q^{(\lambda)}$=$Q^{\rm
  cutoff}$, at which the parton becomes on-the-mass-shell and appears
as a produced hadron. This occurs after $\lambda$ generations of
parton showering.  The generation number $\lambda$ is determined by
\begin{eqnarray}
\lambda = \ln \left ( \frac{Q^{\rm cutoff}}{Q^{\rm jet}} \right ) \biggr / \ln \langle \xi \rangle.
\label{32}
\end{eqnarray}
There is a one-to-one mapping of the initial virtuality $Q^{\rm jet}$
with the initial jet transverse momentum $c_T$ of the evolving parton
as $Q^{\rm jet} (c_T)$ (or conversely $c_T(Q^{\rm jet})$). The cut-off
virtuality $Q^{\rm cutoff}$ maps into a transverse momentum
$c_{T0}$=$c_T(Q^{\rm cutoff})$.  Because of such a mapping, the
decrease in virtuality $Q$ corresponds to a decrease of the
corresponding mapped $ c_T$.  We can infer from Eq.\ (\ref{32}) an
approximate relation between $c_T$ and the number of generations,
$\lambda$,
\begin{eqnarray}
\lambda=\ln \left ( \frac{Q^{\rm cutoff}(c_{T0})}{Q^{(0)}(c_T)}  \right ) \biggr / {\ln  \langle \xi \rangle} \simeq \ln \left ( \frac{c_{T0}} {c_T}  \right ) \biggr / {\ln  \langle \xi \rangle}.~~~~
\end{eqnarray}
Thus, the showering generation number $\lambda$ depends on the
magnitude of the jet momentum $c_T$.  On the other hand,
kinematically, the showering processes degrades the transverse
momentum of the parton $c_T$ to that of the $p_T$ of the produced
hadron as given by Eq.\ (\ref{eq52}), depending on the number of
generations $\lambda$.  The magnitude of the transverse momentum $p_T$
of the produced hadron is related to the transverse momentum $c_T$ of
the parent parton jet by
\begin{eqnarray}
\frac{p_T}{c_T} = \langle z \rangle^{\lambda}=
\langle z \rangle^{{\ln (\frac{c_{T0}}{c_T} )} /{\ln  \langle \xi \rangle}}.
\end{eqnarray}
We can solve the above equation for $p_T$ as a function of $c_T$ and obtain
\begin{eqnarray}
\hspace*{-0.5cm}\frac{p_T}{c_{T0}}
=\left ( \frac{c_T}{c_{T0}} \right )
 ^{1- \mu},
\end{eqnarray}
and conversely
\begin{eqnarray}
\frac{c_T}{c_{T0}}
=\left ( \frac{p_T}{c_{T0}} \right )
 ^{1/(1- \mu)},
\label{23}
\end{eqnarray}
where
\begin{eqnarray}
\mu=\frac{\ln  \langle z \rangle }{\ln \langle \xi \rangle}.
\label{24a}
\end{eqnarray}
In practice $\mu$ (or equivalently, the cut-off parameter $Q^{\rm
  couoff}$ or $c_{T_0}$) is a parameter that can be tuned to fit the
data.  As a result of the virtuality degradation and virtuality
cut-off, the hadron fragment transverse momentum $p_T$ is related to
the parton momentum $c_T$ nonlinearly by an exponent $1-\mu$.

After the showering of the parent parton $c_T$ to
the produced hadron $p_T$, the hard-scattering cross section for the
scattering in terms of hadron momentum $p_T$ becomes
\begin{eqnarray}
&&\hspace*{-1.5cm}
 \frac{d^3\sigma( AB \to p X) }{dy d{\bb p}_T}
= \frac{d^3\sigma( AB \to c X) }{dy d{\bb c}_T}
\frac{d{\bb c}_T}{d{\bb p}_T}.
\end{eqnarray}
Upon substituting the non-linear relation (\ref{23}) between the
parent parton moment $c_T$ and the produced hadron $p_T$ in
Eq.\ (\ref{23}), we get
\begin{eqnarray}
\frac{d{\bb c}_T}{d{\bb p}_T}
=
{\frac{1}{1- \mu}}\left ( \frac{p_T}{c_{T0}} \right )
 ^{\frac{2\mu}{1- \mu}}.
\label{77}
\end{eqnarray}
Therefore under the parton showering from $c$ to $p$, the
hard-scattering invariant cross section $\sigma_{\rm inv}(p_T)$ for
$AB \to p X$ for hadron production becomes
\begin{eqnarray}
&& \hspace*{-0.1cm}  \sigma_{\rm inv}(p_T)=E_c \frac{d^3\sigma( AB\!\! \to\!\! p X) }{dp^3}\biggr |_{y\sim 0}
=\frac{d^3\sigma( AB \to p X) }{dy d{\bb p}_T}\biggr |_{y\sim 0}  \nonumber\\
&&  \!\! \propto
 \frac{\alpha_s^2(Q^2(c_T)) (1-x_{a0}(c_T))^{g_a+\frac{1}{2}}(1-x_{b0}(c_T))^{g_b+\frac{1}{2}}}
{p_T^{n'} \sqrt{1-x_c(c_T)}}, \label{44}
\end{eqnarray}
where
\begin{eqnarray}
n'=\frac{n-2\mu}{1-\mu},~~{\rm with~~} n=4+\frac{1}{2}.
\end{eqnarray}
Thus, the power index $n$ for jet production can be significantly
changed to $n'$ for hadron production because the greater the value of
the parent jet $c_T$, the greater the number of generations $\lambda$
to reach the produced hadron, and the greater is the kinematic energy
degradation.  By a proper tuning of $\mu$, the power index can be
brought to agree with the observed power index $n'$ in hadron production.
The quantity $\mu$ is related to $n'$ and $n$ by
\begin{eqnarray}
\mu= \frac{n'-n}{n'-2}
\end{eqnarray}•
For example, for $\mu$=0.4 one gets $n'$=6.2 and for $\mu=0.6$ one
gets $n'$=8.2.

\section{ Regularization and Further Approximation of 
the Hard-Scattering Integral}

In order to apply the power-law (\ref{44}) to the whole range of $p_T$
for hadron production, we need to regularize it.  Upon choosing the
regularization (\ref{19a}), the differential invariant cross section
$\sigma_{\rm inv} (p_T)$ for the production of a hadron with a
transverse momentum $p_T$ becomes
\begin{eqnarray}
&& \hspace*{-0.3cm}  \sigma_{\rm inv}(p_T)
=\frac{d^3\sigma( AB \to p X) }{dy d{\bb p}_T}\biggr |_{y\sim 0}  \nonumber\\
&&  \!\!\!\!\!\! \propto
 \frac{\alpha_s^2(Q^2(c_T)) (1-x_{a0}(c_T))^{g_a+\frac{1}{2}}(1-x_{b0}(c_T))^{g_b+\frac{1}{2}}}
{(1+ m_T/m_{T0})^{n'} \sqrt{1-x_c(c_T)}}. \label{46}
\end{eqnarray}
In the above equation, the variable $c_T(p_T) $ on the right-hand side
refers to the transverse momentum of the parent jet $c_T$ before
parton showering as given by Eq.\ (\ref{23}),
\begin{eqnarray}
\frac{c_T(p_T)}{c_{T0}} =\left ( \frac{   p_T}{c_{T0}}\right )^{ {(n'-2)}/{(n-2)} }.
\label{eq65}
\end{eqnarray}
The quantities $x_{a0}$, $x_{b0}$, and $x_c$ in Eqs.\ (\ref{44}) are
given by Eq.\ (\ref{5}).

We can simplify further the $p_T$ dependencies of the structure
function in Eq.\ (\ref{46}) and the running coupling constant as
additional power indices in such a way that will facilitate subsequent
phenomenological comparison.  We can cast the hard-scattering integral
Eq.\ (\ref{46}) for hadron production in the nonextensive statistical
mechanical distribution form
\begin{eqnarray}
\hspace*{-0.9cm} \frac{d^3\sigma( AB \to p X) }{dy
d{\bb p}_T}
\biggr |_{y\sim 0}=\sigma_{\rm inv} (p_T)
\sim \frac{A}  {[1+{m_{T}}/{m_{T0}}]^{n}}
, \label{49}
\end{eqnarray}
where
\begin{eqnarray}
n=n'+n_\Delta,
\label{47}
\end{eqnarray}
and $n'$ is the power index after taking into account the
parton showering process, $n_\Delta$ the power index from the
structure function and  the coupling constant.  We  consider the
part of the $p_T$-dependent factor in Eq.\
(\ref{46})
\begin{eqnarray}
f(p_T) =\frac{ \alpha_s^2(c_T(p_T))(1-2
c_T(p_T)/\sqrt{s})^{g_a+g_b+1} }{[1-c_T(p_T)/\sqrt{s}]^{1/2}}
\label{53}
\end{eqnarray}
that is a known function of $p_T$.    We wish to match it to a
nonextensive statistical mechanical distribution with a power
index $n_\Delta$,
\begin{eqnarray}
\tilde f (p_T) = \frac{\tilde
A}{(1+m_T(p_T)/m_{T0})^{n_\Delta}}.
\end{eqnarray}
We match the two functions at two points, $p_{T1}$ and $p_{T2}$ ,
\begin{eqnarray}
f(p_{Ti}) = \tilde f (p_{Ti}), ~~~ i=1,2
\end{eqnarray}
Then we get
\begin{eqnarray}
n_{\Delta}\!=\!\frac{\ln f(p_{T1})   - \ln f(p_{T2})}
{\ln (1+m_T(p_{T2})/m_{T0})\! -\! \ln (1+m_T(p_{T1})/m_{T0})}.~~~~~~
\end{eqnarray}
As $f(p_T)$ is a known function of $p_T$ and $\sqrt{s}$, $n_{\Delta}$
can in principle be determined. The total power index $n$ as given by
(\ref{47}) is also a function of $\sqrt{s}$.

In reaching the above representation of Eq.\ (\ref{49}) for the
invariant cross section for hadrons, we have approximated the
hard-scattering integral $\sigma_{\rm inv} (p_T)$ that may not be
exactly in the form of $A/[1+m_T/m_{T0}]^n$ into such a form. It is
easy then to see that the upon matching $\sigma_{\rm inv}(p_T)$ with
$A/[1+m_T/m_{T0}]^n$ according to some matching criteria, the
hard-scattering integral $\sigma_{\rm inv}(p_T)$ will be in excess of
$A/[1+m_T/m_{T0}]^n$ in some region, and is in deficit in some other
region.  As a consequence, the ratio of the hard-scattering integral
$\sigma_{\rm inv}(p_T)$ to the fitting $A/[1+m_T/m_{T0}]^n$ will
oscillate as a function of $p_T$.  This matching between the physical
hard-scattering outcome that contains all physical effects with the
approximation of Eq.\ (\ref{49}) may be one of the origins of the
oscillations of the experimental fits with the nonextensive
distribution (as can be seen below in Fig.\ 8).

\section{Single-Particle Nonextensive Distribution as a Lowest-Order 
Approximation of the
Hard-scattering Integral }

In the hard-scattering integral Eq.\ (\ref{49}) for hadron invariant
cross section at central rapidity, if we identify
\begin{eqnarray}
n \to \frac{1}{q-1}~~~{\rm and~~}m_{T0} \to \frac{T}{q-1}=nT,
\end{eqnarray}
and consider produced particles to be relativistic so that $m_T$
$\sim$ $ E_T$ $ \sim $ $p_T$, then we will get the nonextensive
distribution of Eq.\ (1) as the lowest-order approximation for the
QCD-based hard-scattering integral.

It is necessary to keep in mind that the outlines leading to
Eqs.\ (\ref{46}) and (\ref{49}) pertains only to average values, as
the stochastic elements and distributions of various quantities have
not been properly taken into account. The convergence of
Eq.\ (\ref{49}) and Eq.\ (1) can be considered from a broader
viewpoint of the reduction of a microscopic description to a single-particle 
statistical-mechanical description. From the microscopic perspective,
the hadron production in a $pp$ collision is a very complicated
process, as evidenced by the complexity of the evolution dynamics in
the evaluation of the $p_T$ spectra in explicit Monte Carlo programs,
for example, in \cite{PYTHIA,HERWIG,ARIADNE}.  There are stochastic
elements in the picking of the degree of inelasticity, in picking the
colliding parton momenta from the parent nucleons, the scattering of
the partons, the showering evolution of scattered partons, the
hadronization of the fragmented partons.  Some of these stochastic
elements cannot be definitive and many different models have been put
forth.  In spite of all these complicated stochastic dynamics, the
final result of Eq.\ (\ref{49}) of the single-particle distribution
can be approximated to depend only on three degrees of freedom, after
all is done, put together, and integrated.  The simplification can be
considered as a ``no hair" reduction from the microscopic description
to nonextensive statistical mechanics in which all the complexities in
the microscopic description ``disappear" and subsumed behind the
stochastic processes and integrations.  In line with statistical
mechanics and in analogy with the Boltzmann-Gibbs distribution, we can
cast the hard-scattering integral in the nonextensive form in the
lowest-order approximation as \cite{CTWW14}\footnote{We are adopting
  the convention of unity for both the Boltzmann constant~$k_B$ and
  the speed of light~$c$.}
\begin{eqnarray}
&&\hspace*{-1.4cm}\frac{d\sigma}{dy d\boldsymbol{p}_T }\biggr |_{y\sim 0} = \frac{1}{2\pi p_T}
\frac{d\sigma}{dy dp_T} \biggr |_{y\sim 0} = A e_q^{-E_T/T} ,
\label{qexponential}
\end{eqnarray}
where
\begin{eqnarray}
&&e_q^{-E_T/T}  \equiv \left[ 1 - \left(1-q\right) E_T/T  \right]^{1/(1-q)},
\nonumber\\
&&e_1^{-E_T/T} =e^{-E_T/T}.\nonumber
\end{eqnarray}
In the above equation,  $E_T$=$\sqrt{m^2+{\bb p}_T^2}$, where $m$ can be taken to be the pion mass $m_\pi$,  and we have assumed
boost-invariance in the region near $y$ $\sim$ 0.  The parameter $q$
is related physically to the power index $n$ of the spectrum, and the
parameter $T$ related to $m_{T0}$ and the average transverse momentum,
and the parameter $A$ related to the multiplicity (per unity rapidity)
after integration over $p_T$.  Given a physically determined invariant
cross section in the log-log plot of the cross section as a function
of the transverse hadron energy as in Fig. 1, the slope at large $p_T$
gives approximately the power index $n$ (and $q$), the average of
$E_T$ is proportional to $T$ (and $m_{T0}$), and the integral over
$p_T$ gives $A$.

\vspace{5mm}
\begin{figure}
\centering
\includegraphics[scale=0.43]{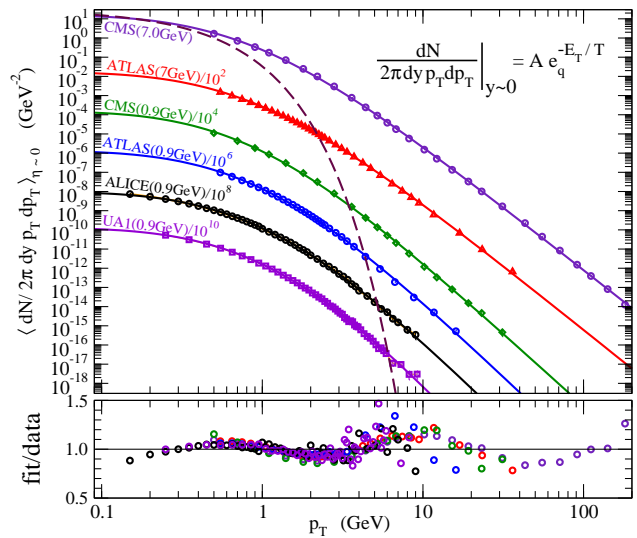}
\caption{(Color online) Comparison of Eq.\ (\ref{qexponential}) with the experimental
  transverse momentum distribution of hadrons in~$pp$ collisions at
  central rapidity $y$.  The corresponding Boltzmann-Gibbs (purely
  exponential) distribution is illustrated as the dashed curve. For a better
  visualization, both the data and the analytical curves have been
  divided by a constant factor as indicated. The ratios data/fit are
  shown at the bottom, where a roughly log-periodic behavior is
  observed on top of the $q$-exponential one. Data are taken from
  \cite{CMS,ATLAS,ALICE,UA1had}.  \label{F4} }
\label{Fig7}
\end{figure}

We can test the above single-particle nonextensive statistical mechanical description
by confronting Eq.\ (\ref{qexponential}) with experimental data.
Fig.\ \ref{Fig7} gives the comparisons of the results from
Eq.\ (\ref{qexponential}) with the experimental $p_T$ spectra at
central rapidity obtained by different Collaborations
\cite{CMS,ATLAS,ALICE,UA1had}.  In these calculations, the parameters of $A$,
$q$ and the corresponding $n$ and $T$ are given in Table III.  The
dashed line (an ordinary exponential of~$E_T$ for $q\to$ 1)
illustrates the large discrepancy if the distribution is described by
Boltzmann-Gibbs distribution.  The results in Fig. 8 shows that
Eq.\ (\ref{qexponential}) adequately describes the hadron $p_T$
spectra at central rapidity in high-energy $pp$ collisions.  We verify
that~$q$ increases slightly with the beam energy, but, for the present
energies, remains always $q\simeq 1.1$, corresponding to a power index
$n$ in the range of 6-10 that decreases as a function of $\sqrt{s}$.

\begin{table}[h]
\centering \caption{Parameters used to obtain fits presented in
  Fig. \ref{F4}.  The values of $A$ is
  in units of {GeV}$^{-2}/c^3$.  } \label{Tab:LC_fits}
\begin{tabular}{|c|c|c|c|c|c|}
\hline\!\!Collaboration \!\!\! &\!$\sqrt{s}$\!& $A$ & $q$ &$n$=1/$(q\!-\!1)$\!&\!T(GeV)\!\\
\hline
CMS~\cite{CMS}&$pp$ at $7$TeV& $16.2$ & $1.151$  & 6.60 & 0.147\\
ATLAS\cite{ATLAS}&$pp$ at $7$TeV& $17.3$ & $1.148$ & 6.73 & 0.150 \\
CMS \cite{CMS}&$pp$ at $0.9$TeV\!& $15.8$ & $1.130$ & 7.65 & 0.128\\
ATLAS\cite{ATLAS}&$pp$ at $0.9$TeV\!& $13.6$ & $1.124$ & 8.09 & 0.140 \\
ALICE\cite{ALICE}&$pp$ at $0.9$TeV\!& $9.95$ & $1.119$ & 8.37 & 0.150 \\
UA1~ \cite{UA1had} &$\bar pp$ at $0.9$TeV\!& 13.1 & 1.109 &  9.21 & 0.154\\
 \hline
\end{tabular}
\end{table}

What interestingly emerges from the analysis of the data in
high-energy~$pp$ collisions is that the good agreement of the present
phenomenological fit extends to the whole~$p_T$ region (or at least
for~$p_T$ greater than $0.2\,\textrm{GeV}/c$, where reliable
experimental data are available)~\cite{Won12}. This is being achieved
with a single nonextensive statistical mechanical distribution with
only three degrees of freedom with data-to-fit ratios oscillating
about unity as in Fig. 8.  Such an agreement suggests that the
nonextensive statistical mechanical distribution may not only be the
phenomenological description of the end product of the parton
showering evolution from jet to hadrons but may have deeper
theoretical significance.  

\section{Summary and Discussions}

Transverse momentum distribution of jets and hadrons provide
complementary and useful pieces of information on the collision
mechanisms and their evolution dynamics. The spectra of jets reveal
the simple hard-scattering production mechanism and they carry the
distinct signature with a power index of $n\sim $ 4 - 5.  On the other
hand, the spectra of hadrons contain additional subsequent dynamics on
the evolution of jets into hadrons but retain the power-law feature of the hard-scattering
process.   Recent description of the
hadron transverse spectrum by a single nonextensive statistical
mechanical distribution leads to the suggestion of  the possible 
dominance of the hard-scattering process, not only in the high $p_T$
region, but also over essentially the whole $p_T$ region, for $pp$ and
$\bar p p$ collisions.
The suggestion represents a synthesizing  description linking
 the simplicity of the whole hadron spetrum  for $pp$ collisions  with
the production  of minijets \cite{Wan91} at $p_T$ of a few GeV and the production of minijets at 
low-$p_T$ \cite{STAR06twopar,Por05,Tra11,Ray11} into 
a single simplifying observation on the dominance of the hard-scattering over the whole $p_T$ region in $pp$ collisions,  with a special  emphasis on the production mechanism.

 We have searched for direct supporting evidences  for the dominance of the
 hard-scattering process in the whole $p_T$ region
 at central rapidity. The first piece of evidence has been found by
 studying the power index for jet production in the lower-$p_T$ region
 in the UA1 and ATLAS  data in high-energy $\bar p p$ and $pp$  collisions, where the
 power index is indeed close to 4 - 5, the signature of pQCD jet
 production. The dominance of the hard-scattering process for the production of
 low-$p_T$ hadron in the central rapidity region is further supported
 by two-particle correlation data where associated particles are
 correlated on the near-side at $(\Delta \phi,\Delta \eta)$$\sim$0,
 with a minimum-$p_T$-biased  or
a high-$p_T$ trigger,
 indicating the production of  angular clusters in
 essentially the whole range of $p_T$.  Additional evidence has been
 provided by the two-particle correlation on the away-side at
 $\Delta\phi \sim \pi$, with a minimum-$p_T$-biased or
a high-$p_T$ trigger, where a produced hadron has been found to
 correlate with a ``ridge" of particles along $\Delta\eta$
 \cite{STAR06,STAR06twopar,Por05,Tra11,Ray11}. The 
 $\Delta\phi\sim \pi$ correlation indicates that the correlated pair
 is related by a collision, and the $\Delta\eta$ correlation in the
 shape of a ridge indicates that the two particles are partons from
 the two nucleons and they carry different fractions of the
 longitudinal momenta of their parents, leading to the ridge
 of~$\Delta \eta$ at $\Delta\phi\sim \pi$.

Hadron production in high-energy $pp$ and $\bar p p$ collisions are
complex processes.  They can be viewed from two different and
complementary perspectives.  On the one hand, there is the successful
microscopic description involving perturbative QCD and nonperturbative
hadronization at the parton level where one describes the detailed
mechanisms of parton-parton hard scattering, parton structure
function, parton fragmentation, parton showering, the running coupling
constant and other QCD processes \cite{Sj87}.  On the other hand, from
the viewpoint of statistical mechanics, the single-particle
distribution may be cast into a form that exhibit all the essential
features of the process with only three degrees of freedom
\cite{Won12,Won13,CTWW14}.  The final result of the process may be
summarized, in the lowest-order approximation, by a power index $n$
which can be represented by a nonextensivity parameter $q$=$(n+1)/n$,
the average transverse momentum $m_{T0}$ which can be represented by
an effective temperature $T$=$m_{T0}/n$, and a constant $A$ that is
related to the multiplicity per unit rapidity when integrated over
$p_T$.  We have 
successfully 
confronted such a phenomenological nonextensive statistical mechanical
description with experimental data.
 We emphasize also that, {\it in
  all cases}, the temperature turns out to be close to the mass of the
pion.

What we may extract from the behavior of the experimental data is that
scenario proposed in \cite{Michael,H} appears to be essentially
correct excepting for the fact that we are not facing thermal
equilibrium but a different type of stationary state, typical of
violation of ergodicity (for a discussion of the kinetic and effective
temperatures see \cite{ET,overdamped}; a very general discussion of the notion of temperature on nonextensive environments
     can be found in \cite{Bir14}).  It should be realized
however that the connection between the power law and the nonextensive
statistical mechanical description we have presented constitutes only
a plausible mathematical outline and an approximate road-map. It will
be of interest in future work to investigate more rigorously the
stochastic parton showering process from a purely statistical
mechanical viewpoint to see how it can indeed lead to a nonextensive
statistical distribution by deductive, physical, and statistical
principles so that the underlying nonextensive parameters can be
determined from basic physical quantities of the collision process.

We can discuss the usefulness of our particle production results in
$pp$ collisions in relation to particle production in AA collisions.
In the lowest approximation with no initial-state and final-state
interactions, an AA collisions at a certain centrality $\bb b$ 
can be considered
as a
collection of binary $N_{\rm bin}({\bb b})$ number of $pp$ collisions.
These binary collisions lead first to the production of primary
particles.  Successive secondary and tertiary collisions between
primary particles lead to additional contributions in a series:
\begin{eqnarray}
&&\hspace*{-0.5cm}\frac{E_pdN^{AA}}{d{\bb p} }({\bb b},{\bb p}) =N_{\rm bin}(\bb b) \frac{E_pdN^{pp}}{d{\bb p} }({\bb p})
\nonumber\\
&+& N_{\rm bin}^2(\bb b) \int d{\bb p}_1 d{\bb p}_2 \frac{dN^{pp}}{d{\bb p}_1 }
 \frac{dN^{pp}}{d{\bb p}_2 }
 \frac{E_p dN ( {\bb p}_1 {\bb p}_2 \to {\bb p} X') }{d{\bb p}}
\nonumber\\
&+& N_{\rm binl}^3(\bb b) \int d{\bb p}_1 d{\bb p}_2d{\bb p}_3 \frac{dN^{pp}}{d{\bb p}_1 }
 \frac{dN^{pp}}{d{\bb p}_2 } \frac{dN^{pp}}{d{\bb p}_3 }
\nonumber\\
&& \hspace*{2.8cm} 
 \times \frac{E_p dN ( {\bb p}_1 {\bb p}_2 {\bb p}_3 \to {\bb p} X') }{d{\bb p}}+....,
\label{58}
\end{eqnarray}
where ${E_p dN ( {\bb p}_1 {\bb p}_2 ... \to {\bb p} X') }/{d{\bb p}}$
is the particle distribution of $\bb p$ after binary collisions of
primary particles $\bb p_1, \bb p_2, ...$.  In addition to the primary
products of a single relativistic hard-scattering ${EdN^{pp}}/{d{\bb
    p} }$ represented by the first term on the right-hand side, the
spectrum in AA collisions contains contributions from secondary and
tertiary products represented by the second and third terms.  In the
next level of approximation, additional initial-state and final-state
interactions will lead to further modifications of the ratio $R_{\rm
  AA}$=$dN^{AA}/[N_{\rm bin}dN^{pp}$] as a function of ${\bb b}$ and
${\bb p}_T$.

The usefulness of our analysis arises from a better understanding of
the plausible reasons why the products from the primary $pp$
scattering can be simply represented by a single nonextensive
statistical mechanical distribution (\ref{qexponential}).  For
peripheral collisions, the first term of Eq.\ (\ref{58}) suffices and
the spectrum of AA collision, normalized per binary collision, would
be very similar to that of the $pp$ collision, as is indeed the case
in Fig.\ 1 of \cite{ALICE13}.  As the number of binary collisions
increases in more central collisions, the second term becomes
important and shows up as an additional component of nonextensive
statistical mechanical distribution with a new set of $n$ and $T$
parameters in the region of low $p_T$, as discussed in
\cite{Urm14,MR15}.

As a concluding remark, we note that the data/fit plot in the bottom
part of Fig.~\ref{F4} exhibit an intriguing rough log-periodicity
oscillations, which suggest corrections to the lowest-order
approximation of Eq.\ (\ref{qexponential}) and some hierarchical
fine-structure in the quark-gluon system where hadrons are
generated. This behavior is possibly an indication of some kind of
fractality in the system. Indeed, the concept of
\emph{self-similarity}, one of the landmarks of fractal structures,
has been used by Hagedorn in his definition of fireball, as was
previously pointed out in \cite{Beck} and found in analysis of jets
produced in~$pp$ collisions at LHC~\cite{GWZW}. This small
oscillations have already been preliminary discussed in Section 8 and
in~\cite{Wilk1,Wilk2}, where the authors were able to mathematically
accommodate these observed oscillations essentially allowing the
index~$q$ in the very same Eq.~\eqref{qexponential} to be a complex
number\footnote{It should be noted here that other alternative to
  complex~$q$ would be log-periodic fluctuating scale parameter~$T$,
  such possibility was discussed in~\cite{Wilk2}.} (see also
Refs.~\cite{logperiodic,Sornette1998}; more details on this
phenomenon, including also discussion of its presence in recent AA
data, can be found in \cite{MR}).\\

\vspace*{0.3cm}
\centerline{\bf Acknowledgements}
\vspace*{0.3cm}

One of the authors (CYW) would like to thank Dr.\ Xin-Nian Wang
for helpful discussions. The research of CYW was supported in part
by the Division of Nuclear Physics, U.S. Department of Energy under Contract DE-AC05-00OR22725, and
the research of GW was supported in part by the National Science
Center (NCN) under contract Nr 2013/08/M/ST2/00598 (Polish
agency). Two of us (L.J.L.C. and C.T.) have benefited from partial
financial support from CNPq, Faperj and Capes (Brazilian
agencies). One of us (CT) acknowledges partial financial support
from the John Templeton Foundation.

\end{document}